\documentclass{mn2e}
\usepackage{graphics}
\usepackage{txfonts} 

\title[Speckle interferometry of CH Cygni]{First spatial resolution of the stellar components of the interacting binary CH Cygni\thanks{Based on observations made with the 6 m BTA telescope, which is operated by the Special Astrophysical Observatory (SAO), Russia}}
	    
\author[J.\ Miko{\l}ajewska et al.]
       {Joanna~Miko{\l}ajewska$^1$\thanks{e-mail: mikolaj@camk.edu.pl},
  Yuri Balega$^2$, Karl-Heinz Hofmann$^3$ 
 and Gerd Weigelt$^3$
        \vspace*{1mm}\\
       $^1$  N. Copernicus Astronomical Center, Bartycka  18, PL-00716 Warsaw, Poland\\
        $^2$ Special Astrophysical Observatory, N.Arkhyz, Karachai-Cherkesia 369167, Russia\\ 
$^3$ Max-Planck-Institut fuer Radioastronomie, Auf dem Hugel 69, 53121 Bonn, Germany}
	     
\date{\fbox{\sc Draft Version}}

\pagerange{\pageref{firstpage}--\pageref{lastpage}}
\pubyear{2008}

\begin{document}

\maketitle

\label{firstpage}

\begin{abstract} We report the first resolved bispectrum speckle interferometry of the symbiotic binary CH Cyg. The measured component separation, $\rho=42 \pm 2$ mas, is consistent with the one derived from the known spectroscopic orbit and distance. In particular, our result implies a total mass of the binary of $M_{\rm t}=M_{\rm g}+M_{\rm wd}=3.7^{+3.5}_{ -1.7}\, \rm M_{\sun}$, which is in good agreement with the value $M_{\rm t}=2.7^{+1.2}_{ -0.6}\, \rm M_{\sun}$ derived from the spectroscopic orbit solution for the red giant and evolutionary contraints. We also show that the radio jets and the bipolar outflow are not orthogonal to the orbital plane of the binary system.
 \end{abstract}


\section{Introduction}
CH Cygni is one of the most fascinating as well as the brightest and closest symbiotic stars. It has been extensively studied from radio through X-rays (e.g. Miko{\l}ajewska, Selvelli  \& Hack 1988; Miko{\l}ajewski, Miko{\l}ajewska \& Khudyakova 1990, 1992;  Crocker et al. 2002; Sokoloski \& Kenyon 2003a, 2003b, Karovska et al. 2007, and references therein). CH Cyg, composed of an M7 giant and a hot companion, most likely an accreting white dwarf, is the record holder for the complexity of variable phenomena found in a single symbiotic object. In particular, the hot component shows very spectacular activity: irregular outbursts accompanied by fast, massive outflows and jets (e.g. Karovska et al. 2007, and references therein). The light curves of CH Cyg (cf. Fig. 3 of Miko{\l}ajewska 2001) show changes with timescales ranging from dozens of years ($\sim 15$-yr orbital period, activity phases, dust obscuration events) through several hundred days (pulsations and rotation of the M giant) down to minutes (flickering observed during active phases). 

Both the light and radial velocity curves of CH Cyg show multiple periodicities. In particular, a $\sim 100$-day photometric period, best visible in the $VRI$ light curves was attributed to the first overtone radial pulsations of the giant (Miko{\l}ajewski et al. 1992), whereas the nature of the long secondary period (LSP) of $\sim 760$ days also present in the radial velocity curve is not clear. Hinkle et al. (1993) proposed a triple model for the CH Cyg system with the symbiotic pair in a 2.1-yr orbit and a G-K dwarf companion in a 15.6-yr orbit which was later replaced by an M giant (Skopal et al. 1996). This triple model, however, turned out to be very controversial, and there are strong arguments that the 2.1-yr period is due to a non-radial pulsation mode of the giant rather than to a close binary, whereas the 15.6-yr velocity variation is the symbiotic orbit (Schmidt et al. 2006). 
Recently, Hinkle, Fekel \& Joyce (2009, hereafter HFJ) refined the orbital elements of CH Cyg and showed that the long period indeed results from the symbiotic binary orbit, and identified the g-mode nonradial pulsation as the leading mechanism for the LSP modulation. Finally, Pedretti et al. (2009) explain the LSP of CH Cyg by the combined effect of non-radial pulsation and  a low-mass companion orbiting the giant.

In this paper we report the first resolved observation of the stellar components of CH Cyg and discuss implications of this result for the geometry of the binary and its complex circumstellar environment.

\section{Observations} 

CH Cyg was observed at several epochs in the frame of speckle interferometric observations of binary and triple star systems carried with the BTA 6\,m telescope of the Special Astrophysical Observatory of the Russian Academy of Sciences. No indication of the secondary in the system was detected until 2004, when the star was successfully resolved for the first time. The October 2004 data were taken under $1.5"$ seeing in two filters with central wavelengths of 545 nm and 600 nm. Our new photon-counting speckle camera based on a 3-stage S25 photocathode intensifier coupled to a fast 1024x1024 px cooled CCD provided significantly higher performance than the detectors used before. Diffraction-limited images of CH Cyg were reconstructed from the two sets of 2000 short-exposure (20 ms) speckle interferograms using the bispectrum speckle interferometry method (Weigelt 1977; Lohmann et al. 1983). The modulus of the object Fourier transform was determined by the classical Labeyrie (1970) method. Speckle interferograms of two unresolved stars,  HD\,180348 and
HD\,182692, were recorded to determine the atmospheric transfer function. Details of the observing procedure and data reduction are given elsewhere (e.g. Balega et al. 2002; Maximov et al. 2003).
The reconstructed CH Cyg images are presented in Fig.1. 
The results obtained for CH Cyg are summarized in Table~\ref{obs}, where the dates of observations are followed by the measured position angle, $\Theta$,  the measured angular separation, $\rho$, the observed magnitude difference, $\Delta m$, as well as the central wavelength, $\lambda$, of the filter used for the observation, and the FWHM of the filter passband, $\Delta \lambda$.
In the case of unresolved observations we list the upper limits for $\rho$. We also note that the maximum magnitude difference in speckle interferometry depends on seeing conditions and the dynamic  range of the detector. For the 2007 observations with the 6 m telescope the limitation was close to 3.5 magnitudes.

\begin{table} 
\caption{Summary of speckle interferometry of CH Cyg} 
\label{obs}
\begin{tabular}{@{}cccccc} 
\hline  
 Date & JD & $\Theta$ & $\rho$ & $\Delta\,m$ & $\lambda/\Delta \lambda$\cr
         &   2\,400\,000+                    & deg  & mas & mag  & nm \cr
\hline 
1982.847 & 45278.8 &  & $<78$ &  & 550\cr
1983.490 & 45513.6 & & $<22$ &  & 525/10 \cr
1984.367 & 45833.6 & & $<22$ & & 600/14 \cr
1984.845 & 46008.2 & & $<22$ & & 600/14 \cr
2004.815 & 53302.2 & 24.1$\pm$2.1 & 43$\pm$2 & 2.03$\pm$0.04 & 545/30 \cr
2004.815 & 53302.2 & 24.6$\pm$3.5 & 41$\pm$3 & 2.20$\pm$0.11 & 600/30 \cr
2007.244 & 54189.5 & & $<19$ & & 550/20 \cr
2007.487 & 54278.5 & & $<19$ & & 550/20 \cr
2007.487 & 54278.5& & $<21$ & & 600/20 \cr
\hline \end{tabular} 
\end{table}

\section{Results and discussion}

\begin{figure}
\resizebox{\hsize}{!}{\includegraphics{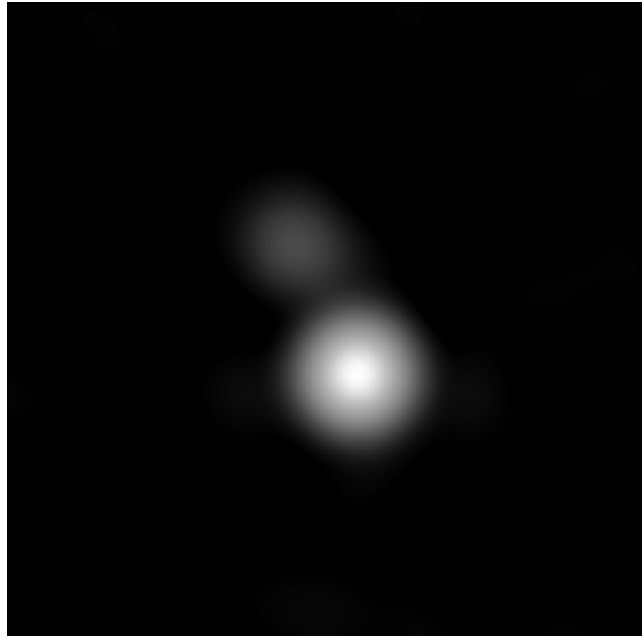}}

\vspace{0.5cm}

\resizebox{\hsize}{!}{\includegraphics{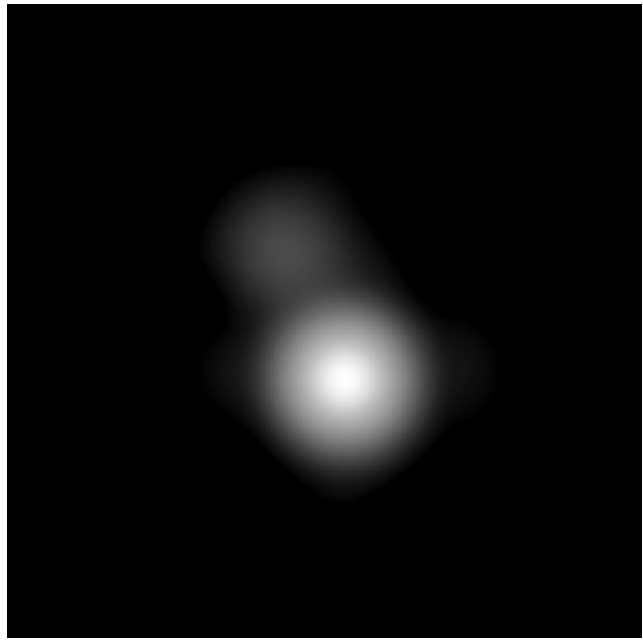}} 
\caption{ Resolved speckle images of CH Cyg obtained through the 545/30 nm (top; separation $43 \pm 2$ mas) and
600/30 nm (bottom) filters, respectively. The field of view in both cases is 0.2 arcsec. North is up and east to the left.} 
\label{images} 
\end{figure}

Relative positions of the CH Cyg components and magnitude differences can be determined from the reconstructed images. The averaged measured separation between the red giant and the $\sim 2$ mag fainter companion visible at the position angle $\Theta=24 \pm 2\degr$ is $42 \pm 2$ mas. This is the only resolved observation of this binary system.
Adopting the revised Hipparcos distance to CH Cyg, $d=244^{+49}_{-35}\, \rm pc$ (van Leeuven 2007), the obtained component separation on the sky corresponds to a real separation  of $10.2^{+2.6}_{-1.9}\, \rm AU$, which is larger than the value quoted by HFJ although overlapping within the error range.
This result immediately implies that the much fainter companion of the red giant cannot be another late M-type giant as suggested by Skopal et al. (1996), and that the binary must be close to their largest separation on the sky, i.e. spectroscopic quadrature. The presence of the second red gaint is also excluded by the interferometric observations made in the infrared in 2004--2006 by Pedretti et al. (2009) who did not find any companion within 50 mas at the better than 100:1 level at H band.

\begin{figure}
\resizebox{\hsize}{!}{\includegraphics{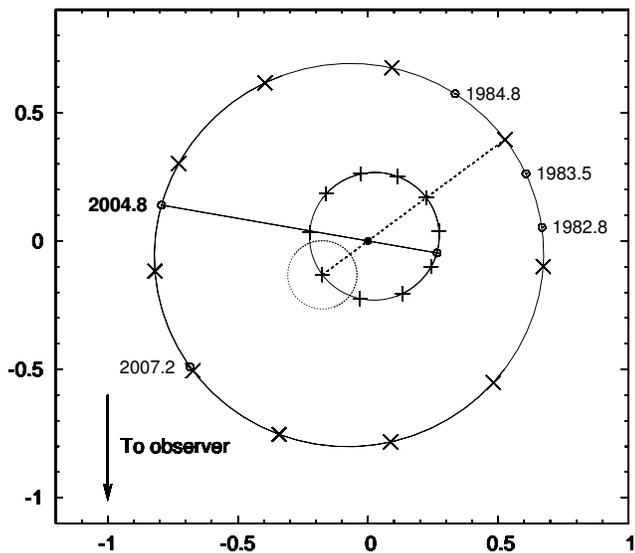}}
      \caption{The orbit of the red giant  (+) and the hot ($\times$)  companion in the CH Cyg binary
system in steps of $\Delta\phi = 0.1$, assuming  a mass ratio of $M_{\rm g}/M_{\rm wd}=3$ (HFJ). In this representation, the stars  move
counter-clockwise. The dotted circle represents the red giant boundary at $\phi = 0$ (periastron passage). The solid dot marks the mass center. Axes are in units of the semi-major axis $a$. The positions of the companion at the times of the speckle interferometric observations are marked by open circles and labeled with dates. The positions of both components during the 2004 resolved observation are connected by a solid line whereas dotted line shows the component position during periastron.}
         \label{orbit}
   \end{figure}

Since the spectroscopic orbit and the inclination of the binary orbit are known, the measured separation can be compared to the calculated separation on the sky using the orbital elements from Table 2 of HFJ and assuming $i \sim 90\degr$ (the system is eclipsing). The results of such calculations are presented in Fig.~\ref{orbit} and in Table~\ref{model} (Set 1, 3rd column). It is clear that in 2004, the binary components of CH Cyg were indeed very close to the quadrature and approaching the apoastron. Combining the calculated $\rho_{\rm cal}  = 1.06\,a$ with the observed value of 42 mas results in a semi-major axis of the true orbit of $a=39.6$ mas, which corresponds to $9.7^{+2.4}_{-1.8}[d/244\, \rm pc]$ AU, and a total mass, $M_{\rm g}+M_{\rm wd}=3.7^{+3.5}_{ -1.7}\, \rm M_{\sun}$, with the errors set mostly by the uncertainty of the distance.  
These values are in good agreement with the binary separation, $a=8.7^{+1.1}_{-0.7}$ AU, and the masses $M_{\rm g}=2^{+1}_{-0.5}\, \rm M_{\sun}$ and  $M_{\rm wd}=0.70^{+0.22}_{-0.09}\, \rm M_{\sun}$ of the red giant and the white dwarf, respectively,  derived from the spectroscopic orbit solution for the red giant and evolutionary contraints  (HFJ). 
Our semi-major axis of the true orbit combined with the binary separation resulting from the spectroscopic orbit solution can also be used to constrain  
a geometrical distance to CH Cyg. Our value of $39.6$ mas combined with 8.7 AU from HFJ implies  $d=220^{+40}_{ -28}$  pc. 

\begin{figure}
\resizebox{\hsize}{!}{\includegraphics{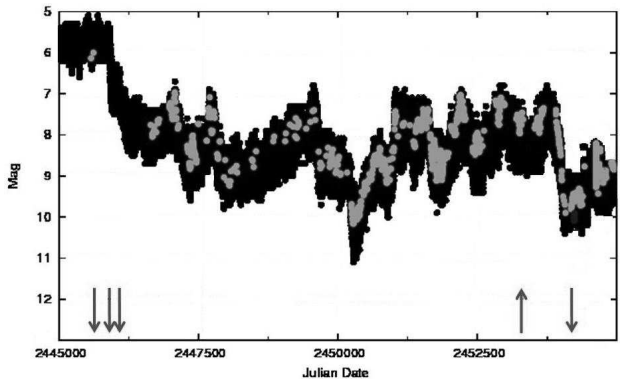}}
      \caption{The AAVSO $m_{\rm vis}/V$ (black and gray dots, respectively) light curve for CH Cyg with our speckle observations marked by vertical arrows pointing downwards (unresolved) and upwards (resolved), respectively.}
         \label{lc}
   \end{figure}

The 2-magnitude difference between the visual optical brightness of the red giant and its companion implies an absolute magnitude of the companion of $M_{\rm opt, h}\sim 2$--3 mag ($L_{\rm opt, h} \sim 5$--$10\, \rm L_{\sun}$) which is reasonable for the mild activity stage of the hot component (e.g. Miko{\l}ajewska et al. 1988). In 2004, the optical spectra indeed showed remarkable H\,{\sc I}, Fe\,{\sc II} and [O\,{\sc III}] line emission, likely originating in an accretion disk (Yoo 2007),  whereas the AAVSO light curve (Fig.~\ref{lc}) shows that CH Cyg was at a similar brightness level as in 1985$-$86, when the optical luminosity of hot component was a few L$_{\sun}$.

In 2004--2006,  Pedretti et al. (2009) performed interferometric observations of CH Cyg using the infrared optical telescope array (IOTA) and the Keck-1 telescope with aperture masking. Combining their null-detection of a companion at the better than 100:1 level at H-band ($\Delta H= H_{\rm h}$--$H_{\rm g} \ga 5$ mag) with our optical detection ($\Delta m_{\rm opt} \sim \Delta V \sim 2$ mag) we estimate the near infrared colour of the companion  relative to the red giant of $(V$--$H)_{\rm h}$--$(V$--$H)_{\rm g} \la$--3. The optical and near infrared light curves of CH Cyg (e.g. Miko{\l}ajewski et al. 1992) indicate  $V$--$H \sim 7.5$ in 1986 in agreement with slightly reddened late type M6--M7\, III giant (e.g. Straizys 1992), and  $(V$--$H)_{\rm h} \la 4.5$.  On the other hand, analyses of the ultraviolet and optical spectra of CH Cyg (e.g. Miko{\l}ajewska et al. 1988) showed that the companion spectrum is a combination of highly variable blue B/F stellar-type continuum and bf+ff emission of ionized gas, for which we estimate $(V$--$H)_{\rm h} \la 1$--2. So, the null-detection in the infrared  is entirely consistent with the expected blue spectral energy distribution of a symbiotic companion detected in the optical light by our speckle inteferometry. 

The predicted orbital positions in the 3rd column of Table~\ref{model}
and Fig.~\ref{orbit} suggest that the binary components should also be resolvable at other epochs. In particular, the separation should be $\sim 32$ mas in 1983, and $\sim 35$ in 2007. 
The fact that CH Cyg was resolved in 2004 and unresolved in 1983, and 2007 can be understood if the separation was in fact lower than our predicted values which are uncertain due to the errors of the adopted orbital elements and/or if the component magnitude difference was larger than in 2004.
In 2007 the visual magnitude of CH Cyg dropped to $V \sim 10$ mag, and has remained at this very low level till 2008 (Fig.~\ref{lc}). Such a low optical state was observed before only in 1996$-$97 when also no signs of the hot companion were observed (e.g. Sokoloski \& Kenyon 2003). So, the lack of detection of the giant companion in 2007 may be due to its very low brightness. However, in 1983, CH Cyg was in the high activity stage, and the two components should have had comparable optical brightness. So, the first possibility seems more likely. In fact, Fig.~\ref{orbit} suggests that 
the projected binary separation would remain in 2004 practically the same, and would decrease for 1983$-$84 and 2007  if the periastron would occur somewhat earlier and/or the $\omega$ value would be somewhat larger. 
Columns 4$-$6 of Table~\ref{model} give the projected component separations for 3 slightly modified sets of orbital parameters:

\begin{itemize}

\item Set 2: the orbital elements from HFJ but $T_0=2\,4445489$, i.e. decreased by its 1-$\sigma$ value; 

\item Set 3: the orbital elements from HFJ but $\omega=230$, i.e. increased by its 1-$\sigma$ value; 

\item Set 4: the orbital elements from HFJ but $T_0=2\,4445489$, and $\omega=230$.
\end{itemize}

These numbers show that particularly the last set of modified orbital parameters excellently reproduces the resolved and unresolved speckle observations of CH Cyg. The semi-major axis of the true orbit, $a\sim40$ mas for all orbital parameter sets in Table~\ref{model}. We also note that varying the eccentricity of HFJ ($e=0.122 \pm 0.024$) by its $\pm 1$-$\sigma$ value does not change much the $\rho_{\rm cal}$ value.

\begin{table} 
\caption{Calculated separations of the CH Cyg binary components projected on the sky in units of the                                                       semi-major axis of true orbit, $a$.} 
\label{model}
\begin{tabular}{@{}cccccc} 
\hline  
\multicolumn{1}{c}{Date} & \multicolumn{1}{c}{JD} & \multicolumn{4}{c}{$\rho_{\rm cal}/a$}  \cr
         &   2\,400\,000+       & Set  1(HFJ) & Set 2 & Set 3 & Set 4 \cr
\hline 
1983.490 & 45513.6 &  0.81 & 0.68 & 0.71 & 0.55 \cr
1984.367 & 45833.6 &  0.57 & 0.38 & 0.41 & 0.19 \cr
1984.845 & 46008.2 & 0.40 & 0.18 & 0.21 & 0.02 \cr
2004.815 & 53302.2 & 1.06 & 1.09 & 1.07 & 1.06 \cr
2007.244 & 54189.5 & 0.91 & 0.79 & 0.74 & 0.59 \cr
2007.487 & 54278.5 & 0.86 & 0.72 & 0.68 & 0.52 \cr
\hline \end{tabular}
\end{table}

The very high orbit inclinaton, $i \sim 90\degr$ (Miko{\l}ajewski et al. 1987; Solf 1987; HFJ) implies a position  angle of the line of nodes -- the binary orientation on sky, of $\Omega  \sim  \Theta=24\degr$. 

\begin{figure}
\resizebox{\hsize}{!}{\includegraphics{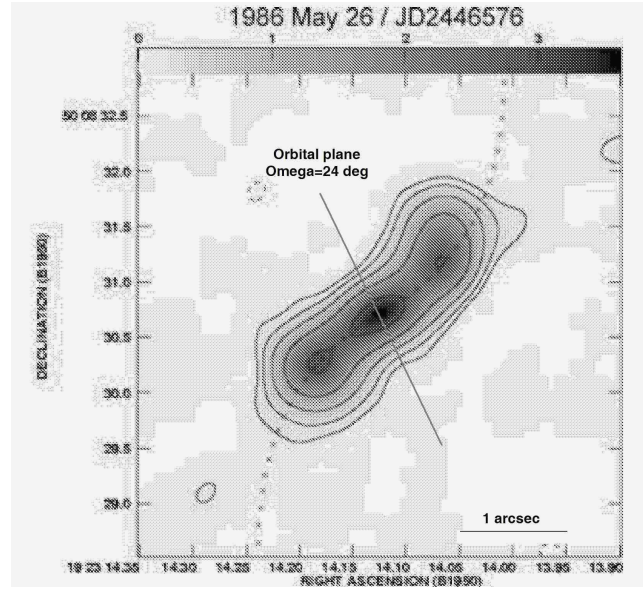}}
      \caption{VLA 5-GHz radio image obtained in May 1986  (from Crocker et al. 2002) showing the jets formed in 1984 with marked position of the binary orbital plane.}
         \label{nebula}
   \end{figure}

The knowledge of $\Omega$ provides unprecedented help to study the very complex morphology of the circumstellar environment of CH Cyg.
First resolved observations of CH Cyg obtained with VLA in mid-1984 and continued in 1985$-$86 revealed two-sided radio jets with an angular expansion rate of 1.1 arcsec/yr and a position angle of $\sim 135\degr$ (Taylor, Seaquist \& Mattei 1986; Taylor 1988). A very compact optical nebulosity (in [O\, {\sc III}]5007) located about 1 arcsec NW of the star was also detected in September 1986 (Solf 1987).  These fast jets were most likely formed as the result of disruption of the inner accretion disc (Sokoloski \& Kenyon 2003a,b).
A decade later, ground-based [O\,{\sc II}] and [N\,{\sc II}] and HST images obtained in 1996$-$99 by Corradi et al. (2001) revealed a very complex, ionized nebula extending out to 18 arcsec. They also found evidence for a bipolar outflow with an extension and position angle consistent with those of the radio jet formed in 1984 and extrapolated to the time of their observations.
Multiple component structure is also visible in a Chandra X-ray image obtained in 2001, which shows a loop-like structure to the south and extension towards the northwest at $\sim 330\degr$ (Karovska et al. 2007).
The comparison of the orientation of the binary on sky with the orientation of the jets as well as other elongated structures in the nebula suggests that these jets and any other outflows are not exactly perpendicular to the orbital plane. In particular, the fast radio jets were inclined at $\sim 70\degr$ to the orbital plane as shown in Fig.~\ref{nebula}. 

Although it is usually assumed that collimated outflows from interacting binaries occur along the polar axis of the orbit, there is not much direct observational evidence for this. Therefore, high spatial resolution observations of the stellar components of such binaries provide important insights on mechanisms powering jets and other outflows from these systems.
CH Cyg is the second symbiotic system -- and in general interacting binary -- showing jets and bipolar outflows with the stellar components resolved on sky. The first one was another symbiotic star, R Aqr, resolved by 
VLA observations at 7 mm, which showed line-less continuum emission associated with the hot companion/accretion disk and a  $v = 1$, $J = 1$--0, SiO maser emission source associated with the long-period--variable (LPV) envelope separated by about 55 mas (Hollis, Pedelty, \& Lyon 1997). This VLA astrometry is in excellent agreement with the recently obtained spectroscopic orbit solution (Gromadzki \& Miko{\l}ajewska 2009).

Interestingly, both CH Cyg and R Aqr seem to contain a magnetic white dwarf as suggested by  periodic components in their flickering light curves (Miko{\l}ajewski et al. 1990; Nichols et al. 2007), and their activity is powered by unstable accretion. Depending on the mechanism involved in the jet production, the ejection should presumably occur either along the magnetic axis or the polar axis of the accretion disc. Note that both axes can, in principle, be more or less inclined to the orbital plane. 
In R Aqr the radio/X-ray jets are practically perpendicular to the binary orbit, whereas the bipolar nebula is not. 
In the case of CH Cyg,  the symetry axis of both the radio jets as well as that of the inner ($\sim 6$\, arcsec across) nebula are inclined to the orbital plane. 
To explain the radio observations collected over 16 years (1985-2001), Crocker et al. (2002) proposed a precessing radio jet, with a precession opening angle of 35$\degr$ and a position angle (on the sky) of the precession axis of $140\degr$. The period of this precession, $\sim 6519$ days, 
is somewhat ($\sim 20$--$30$\, \%) 
longer than the orbital period. The mechanism causing the precession is not clear, although Crocker et al. favored warping of the collimating accretion disc by a magnetic white dwarf. 
We also note that the hypothetical close companion orbiting the red giant (Pedretti et al. 2009) has little if any influence on the jet formation and geometry unless the jets originate from the red giant which in our opinion is very unlikely. 
   
Finally, there are also several symbiotic systems with spatially resolved nebulae, and for some of them, the orientation of the binary orbit on sky can be derived from spectropolarimetric studies, which also allows one to distinguish between the polar and equatorial outflows. The scattering geometry and nebular stuctures are aligned in all these systems although the bipolar outflows may be not orthogonal to the orbital plane in all cases (e.g. Miko{\l}ajewska 1999).
Of particular interest are the other active systems with occasional jet ejections following optical/UV outbursts. For example, resolved radio emission practically aligned with the binary axis was detected in AG Dra (Miko{\l}ajewska 2000), and Z And (Brockshop et al. 2004). In both systems, this emission was associated with the hot component outbursts and presumably originated from jets ejected from the binary systems. Although the interpretation of the multiple outburst behaviour of symbiotic stars is not as straightforward as in the case of the accretion-powered CH Cyg and R Aqr, the hot component of Z And also appears to be a magnetic white dwarf, and the outbursts are probably due to some accretion disc instability (Sokoloski et al. 2006). The main difference between Z And-like outbursts and the activity of CH Cyg, and R Aqr is that the hot component in the former burns the accreted hydrogen more or less stably whereas the latter do not.

Subsequent monitoring to determine precisely the astrometric binary orbit and multifrequency imaging of the innermost ($\la 1$ arcsec) nebula of CH Cyg is necessary to get better constraints on the geometry and mechanisms of the complex outflows from this interacting binary system.

\section{Conclusions}

We have succesfully resolved the stellar components of the symbiotic binary CH Cyg with bispectrum speckle interferometry. The measured component separation on sky, $\rho=42 \pm 2$ mas, is consistent with the one derived from the known spectroscopic orbit and the Hipparcos distance. In particular, our result implies a total mass of the binary of $M_{\rm t}=M_{\rm g}+M_{\rm wd}=3.7^{+3.5}_{ -1.7}\, \rm M_{\sun}$, which is in good agreement with the value $M_{\rm t}=2.7^{+1.2}_{ -0.6}\, \rm M_{\sun}$ derived from the spectroscopic orbit solution for the red giant and evolutionary contraints by HFJ.
The 2-mag difference between the visual optical brightness of the red giant and its companion is reasonable for CH Cyg's mild activity stage indicated by the AAVSO light curve and remarkable HI, FeII, and [OIII] line emission detected in 2004. Such a symbiotic blue companion is also consistent with the null-detection in infrared reported by Pedretti et al. (2009).

Given the very high orbit inclination of CH Cyg, the measured position angle, $\Theta=24 \pm 2$, is very close to the position angle of the line of nodes - i.e. the binary orientation on sky. This means that the  collimated outflows from the binary occur along the polar direction, although the symmetry axis of the inner nebula and radio jets are not exactly orthogonal to the orbital plane.

Subsequent monitoring program to determine precisely the astrometric binary orbit and provide better constraints on the component masses, the geometry of the binary as well as the geometric distance is ongoing.

\subsection*{ACKNOWLEDGEMENTS} 
We acknowledge with thanks the variable star observations from the AAVSO International Database contributed by observers worldwide and used in this research. We thank M. Gromadzki for preparing Fig.~\ref{orbit}.
This study was partly supported by the 
Polish Research Grant No. N203\,395534.

\label{lastpage}

\end{document}